\documentstyle[12pt,epsfig]{article}
\title{\hspace{10cm} {\large Budker INP 97--47\\
\hspace{10cm}May 1997} \\
\vspace{0.5cm} 
 {\bf Ultimate  luminosities and energies of photon colliders}
\footnote{Talk at the Int.Symp. on 
Future High Energy Colliders, ITP, UCSB, Santa Barbara, October
21--25, 1996}}
\author{Valery Telnov \\
 \it{Institute of Nuclear Physics,
630090, Novosibirsk, Russia}} 
\date{}
\begin{document}
\newcommand{\EP}{\mbox{e$^+$}}
\newcommand{\EM}{\mbox{e$^-$}}
\newcommand{\EPEM}{\mbox{e$^+$e$^-$}}
\newcommand{\EMEM}{\mbox{e$^-$e$^-$}}
\newcommand{\GG}{\mbox{$\gamma\gamma$}}
\newcommand{\GE}{\mbox{$\gamma$e}}
\newcommand{\TEV}{\mbox{TeV}}
\newcommand{\GEV}{\mbox{GeV}}
\newcommand{\LGG}{\mbox{$L_{\gamma\gamma}$}}
\newcommand{\EV}{\mbox{eV}}
\newcommand{\CM}{\mbox{cm}}
\newcommand{\MM}{\mbox{mm}}
\newcommand{\NM}{\mbox{nm}}
\newcommand{\MKM}{\mbox{$\mu$m}}
\newcommand{\SEC}{\mbox{s}}
\newcommand{\CMS}{\mbox{cm$^{-2}$s$^{-1}$}}
\newcommand{\MRAD}{\mbox{mrad}}
\newcommand{\IND}{\hspace*{\parindent}}
\newcommand{\E}{\mbox{$\epsilon$}}
\newcommand{\EN}{\mbox{$\epsilon_n$}}
\newcommand{\EI}{\mbox{$\epsilon_i$}}
\newcommand{\ENI}{\mbox{$\epsilon_{ni}$}}
\newcommand{\ENX}{\mbox{$\epsilon_{nx}$}}
\newcommand{\ENY}{\mbox{$\epsilon_{ny}$}}
\newcommand{\EX}{\mbox{$\epsilon_x$}}
\newcommand{\EY}{\mbox{$\epsilon_y$}}
\newcommand{\BI}{\mbox{$\beta_i$}}
\newcommand{\BX}{\mbox{$\beta_x$}}
\newcommand{\BY}{\mbox{$\beta_y$}}
\newcommand{\SX}{\mbox{$\sigma_x$}}
\newcommand{\SY}{\mbox{$\sigma_y$}}
\newcommand{\SZ}{\mbox{$\sigma_z$}}
\newcommand{\SI}{\mbox{$\sigma_i$}}
\newcommand{\SIP}{\mbox{$\sigma_i^{\prime}$}}
\maketitle
\begin{abstract}
  A  photon collider luminosity and its  energy are determined by the
parameters of an electron-electron linear collider (energy, power,
beam emittances) and collision effects. The main collision effect is the
coherent \EPEM\ pair creation. At low energies (2E$<$ 0.5--1 TeV) this
process is suppressed due to repulsion of electron beams. In this
region $\LGG(z>0.65) \ge 10^{35} \CMS\ $ is possible
($10^{33}-10^{34}$ is sufficient). At higher energies the limited
average beam power and coherent pair creation restrict the maximum
energy of photon colliders (with  sufficient luminosity) at
$E_{cm}\sim$ 5 TeV. Obtaining high luminosities requires the development
of new methods of production beams with low emittances such as a laser
cooling.

\end{abstract}

\section{Introduction}

Linear colliders offer the unique opportunities to study \GG, \GE\
interactions. Using the laser backscattering method one can obtain
\GG\ and \GE\ colliding beams with an energy and luminosity comparable
to that in \EPEM\ collisions  or even higher (due to the absence
of some beam collision effects). This can be done with a relatively small
incremental cost. The expected physics in these collisions 
 is very rich and complementary to that in \EPEM\
collisions. Some characteristic examples are:

\begin{itemize}
\item a \GG\ collider provides the  unique opportunities to measure the
  two-photon decay width of the Higgs boson, and to search for
  relatively heavy Higgs states in the extended Higgs models such as
  MSSM;

\item a \GG\ collider is an outstanding $W$ factory, with a $WW$ pair
  production cross section by a factor of 10--20 larger than that in
  \EPEM\, and with a potential of producing $10^6-10^7$ $W$'s per
  year, allowing a precision study of the anomalous gauge boson
  interactions;

\item a \GG, \GE\ collider is a remarkable tool for searching for new
charged particles, such as supersymmetric particles, leptoquarks,
excited  states of electrons, etc., as in \GG, \GE\ collisions they are
produced with cross sections larger than those in \EPEM\ collisions;

\item Charged sypersymmetric particles with massws higher than the
beam energy could be produced with a \GE\ collider.
\end{itemize}

\begin{figure}[!hbp]
\centering
\epsfig{file=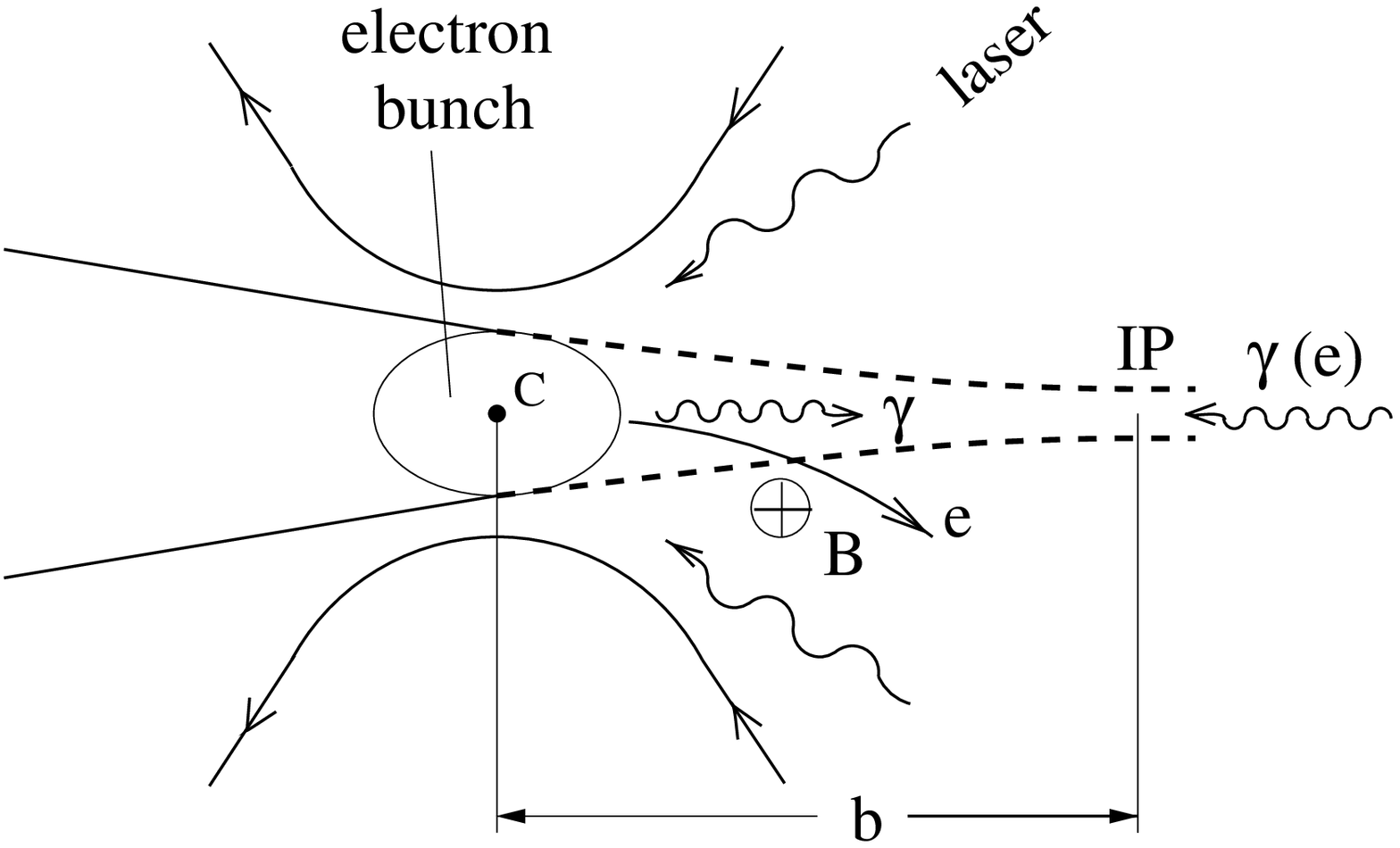, height=8.5cm,angle=-90}
\caption{Scheme of  \GG; \GE\ collider.}
\label{gensch}
\end{figure}

The general scheme of a photon collider is shown in Fig.~\ref{gensch}.

Two electron beams after the final focus system are traveling
toward the interaction point (IP).  At a  distance of about 1 cm
upstream from the IP, at a conversion point
(CP), the laser beam is focused and Compton backscattered by the
electrons, resulting in the high energy beam of photons.
 With reasonable laser
parameters one can ``convert'' most of electrons to high energy
photons. The photon beam follows the original electron direction
of motion with a small angular spread
of order $1/\gamma $, arriving at the IP in a tight focus, where it
collides with a similar opposing high energy photon beam or with an
electron beam. The photon spot size at the IP may be almost
equal to that of electrons at IP and therefore the luminosity of
\GG, \GE\ collisions will be of the same
order as the ``geometric'' luminosity of basic $ee$ beams.

\begin{figure}[!hb]
\centering
\epsfig{file=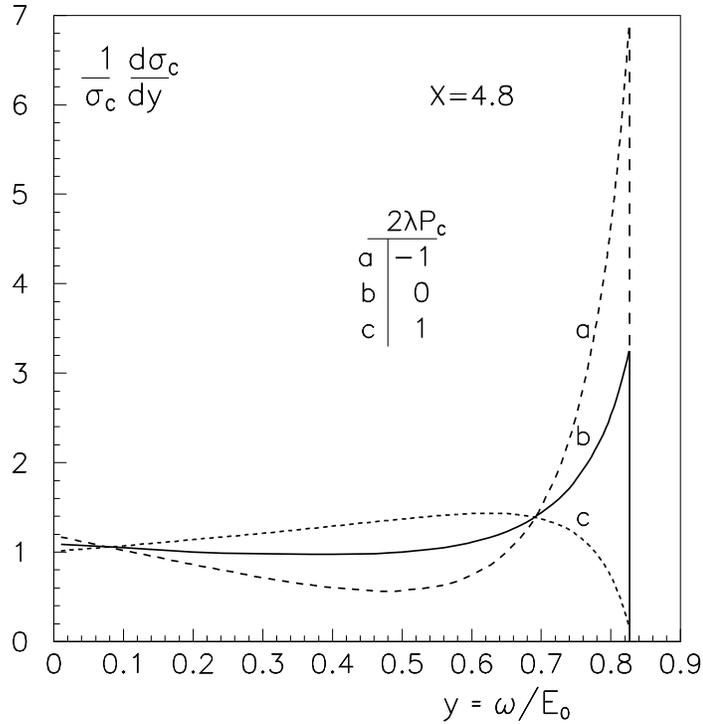, height=12.5cm,angle=0}
\caption{Spectrum of the Compton scattered photons for different
polarizations of laser and electron beams}.
\label{fig4}
\end{figure}

The energy spectrum of photons after the Compton scattering for various
polarization of electrons and laser photons is shown in
fig.~\ref{fig4}. At the optimum laser wave length (below the threshold
of \EPEM\ pair creation), the maximum energy of scattered photons is about
82\% of the initial electron energy. Photons in the high energy part
of spectrum can have high degree of polarization. This part of
spectrum is most valuable for experiment. Below we will deal mainly
with \GG\ luminosity produced by high energy photons.

Here we will not consider the general features of photon colliders, they
are discribed in papers\cite{GKST81}--\cite{TEL95} and Proceedings of
the Berkeley Workshop~\cite{BERK}.

Preliminary studies show that the photon colliders with an energy
2E$\sim$500 GeV and acceptable luminosity \LGG\ $\sim 10^{33}\CMS$ (at
$z =W_{\gamma\gamma}/2E>0.7$ ) can be
built~\cite{BERK,LOW,NLC}. However, physicists will be happy to have
larger luminosity to study details of some processes, for example
$\gamma\gamma \to Higgs$. It is also of interest to investigate main
properties of the photon colliders at higher energy. We know that \EPEM\
linear colliders at an energy above 1--2 TeV have serious
problems. Is it easier to explore this region with a photon collider?

The present paper is focused mainly on the study of limitations on the
energy and luminosity of photon colliders.  We understand that linear
colliders will be built not sooner than in one decade and will work
another one--two decades. Therefore in discussions we will not
confine ourselves by the present achievements and technologies. This
concerns laser parameters (for high energy photon collider a free
electron laser is needed) and emittances of electron beams.  Although
FELs with close parameters do not exist, but there are projects of
such lasers. The way to very low emittances (laser cooling) was
discussed at the first ITP symposium~\cite{TCOOL}.
  
 In the later paper it was also found how to overcome the problem of
nonlinear effects in the conversion region which leads to the linear
growth of the laser flash energy with the  increase in the electron beam
energy. In the proposed method of `stretching' the conversion region
the required laser flash energy does not grow at all with energy.

\section{Requirements to \GG\ luminosity}

 Cross sections of the charged particle production in \GG\ collisions
are somewhat higher than those in \EPEM\ collisions. At $E > Mc^2$ 
the ratio of cross sections are the following: \\ 
   $\sigma_{\GG\to H^+H^-}/\sigma_{\EPEM\to H^+H^-} \sim 4.5;\\
\vspace*{2mm}
\hspace*{3cm}\sigma_{\GG\to t\bar{t}}/ \sigma_{\EPEM\to t\bar{t}} \sim 4; \\
\vspace*{2mm}
\hspace*{3cm}   \sigma_{\GG\to W^+W^-}/\sigma_{\EPEM\to W^+W^-} (|cos
   \vartheta|<0.8) \sim 15; \\
\vspace*{2mm}
\hspace*{3cm} \sigma_{\GG\ \to \mu^+\mu^-}/ \sigma_{\EPEM\to
    \mu^+\mu^-} (|cos\vartheta|<0.8)\sim8.5$.

To have the same statistics (but complementary physics) in \GG\
collisions the luminosity can be smaller than that in \EPEM\
collisions by a factor of 5.

  Cross sections decrease usually as $1/S\; (S=E_{cm}^2)$,
therefore the luminosity should grow proportionally to S. A reasonable
scaling for the required \GG\ luminosity (in the high energy peak of the
luminosity distribution) at \GG\ collider is
\begin{equation}
\LGG\ \sim 3\cdot10^{33}S,\;\CMS.
\end{equation}
With such a luminosity for the time $t=10^7\;c$ one can detect \\ 
\vspace*{2mm}
\hspace*{3cm}$3.5\cdot10^3\;H^+H^-,\;\; \\
\vspace*{2mm}
\hspace*{3cm}2\cdot10^4 \;\mu^+\mu^-(|cos\vartheta|<0.8);\;\; \\
\vspace*{2mm}
\hspace*{3cm}2\cdot10^4 \;t\bar{t};\;\; \\
\vspace*{2mm}
\hspace*{3cm}2\cdot10^5 \;W^+W^-(|cos\vartheta|<0.8);\;\; \\
\vspace*{2mm}
\hspace*{3cm}2\cdot10^6\cdot S(TeV^2) \;W^+W^-$.

 Somewhat larger luminosity ($\sim 10^{33}$) is required for the  search
 and study of the `intermediate' ($M_H \sim 100-200\;\GEV$) Higgs
 boson which is produced as a single resonance in \GG\ collisions. We
 will see that such a level of luminosities at low energies is not a
 problem. The scaling (1) will be used for estimation of the maximum energy
  of photon colliders.

   Other important problem at high luminosities is a background due to
large total cross section $\sigma_{\GG\to hadrons}\sim
5\cdot10^{-31}\;\CMS.$ It consists of particles with $P_t\sim 0.5$ GeV
uniformly (at large angles) distributed over the pseudorapidity
$\eta=-ln\;tan(\vartheta/2)$ with $dN/d\eta \sim 7$ at $2E=500$
GeV. Particle density grows only logarithmically with  energy.

 The average number of hadron events/per bunch crossing is about one
at $\LGG(z>0.65)=10^{34}\;\CMS$ and at the typical collision rate 5 kHz.
In this paper, we are interested mainly by the luminosity in the high
energy part of luminosity spectrum.  However, in the scheme without
deflection of used electron beams the total \GG\ luminosity is larger
than the `useful' $\LGG(z>0.65)$ by a factor 5--10 due to collisions of low
energy Compton photons and beamstrahlung photons.  This low energy
collisions increase background by a factor 2--3. At $E_{cm}=5$ TeV with
required $\LGG(z>0.65)\sim 10^{35}$, this leads to about 30
(effectively) high energy $\GG\to hadron$ events per bunch
crossing. Similar number of hadronic events/collision is expected at
LHC. However, there is important an difference between pp and \GG\
colliders: in the case of an interesting event (high $P_t$ jets and
leptons) the total energy of final products at photon colliders is
equal to  $E_{cm}$, while at proton colliders it is only about
(1/6)$E_{cm}$. The ratio of the signal to background at photon
colliders is better by a factor 6 at the same number of hadronic
events per crossing. Moreover, during the reconstruction of an interesting
event one can subtract smooth hadronic background and only its
fluctuations are important which are proportional to $\sqrt{L}$. It means
that for $L\propto E^2$ (and fixed collision rate) the ratio of signal
to background is almost constant (decreases only logarithmically). 

  The above arguments shows that the problem of hadronic background is
not dramatic for photon colliders. Of course, some increase in the 
collision rate with an increase in the luminosity will be useful.  We
will see that due to collision effects the optimum number of particles
in one bunch should decrease with  energy, that naturally
leads to an increase in the collision rate.

\section{Collision effects. Coherent pair creation.}    

There are two basic collision schemes \cite{TEL95}:

\underline{Scheme A (``without deflection'').} There is no magnetic
deflection of the spent electrons and all particles after the conversion
region travel to the IP. The conversion point may
be situated very close to the IP.

\underline{Scheme B (``with deflection'').} After the conversion region
particles pass through a region with a transverse magnetic field where
electrons are swept aside. Thereby, one can achieve
a more or less pure \GG\ or \GE\ collisions.

  During beam collision, photons are influenced by the field of the
opposing electron beam. One of the important processes in this field
is a conversion of photons into \EPEM\ pairs (coherent pair
creation)\cite{CHEN}. Under a certain conditions, the conversion length
is shorter than the length of interaction region ($\sim \sigma_z$) and
\GG\ luminosity is suppressed.

\IND\  The probability of pair creation per unit length  by  a photon
with the energy $\omega$ in the magnetic field $B (|B|+|E|)$  for our
case) is \cite{CHEN},\cite{TEL90},.

\begin{equation}
\mu (\kappa) = \frac{\alpha^2}{r_e} \frac{B}{B_0} T(\kappa),
\end{equation}
where $\kappa = \frac{\omega}{mc^2}\frac{B}{B_0},\;\; B_0=
\frac{m^2c^3}{e \hbar} = \frac{\alpha e}{r_e^2} = 4.4 \cdot 10^{13}\;
G$ is the the critical field, $r_e = e^2/mc^2$ is the classical radius
of electron.

\begin{equation}
T(\kappa) \approx 0.16 \kappa^{-1} K^{2}_{1/3}(4/3\kappa)
\end{equation}
\hspace*{45mm} $\approx 0.23 \exp (-8/3\kappa)\hspace{15mm} \kappa < 1 $ \\
\vspace{2mm}
\hspace*{45mm} $\approx 0.1 \hspace{31mm} \kappa = 3-100  $ \\
\vspace{2mm}
\hspace*{45mm} $ \approx 0.38 \kappa^{-1/3} \hspace{27mm}
\kappa >100$ 

In our case, $\omega \sim  E_0$ ,  therefore one can put
\begin{equation}
\kappa \sim \Upsilon \equiv \gamma B/B_0\;\;.
\end{equation}
The probability  to create \EPEM\ pair during the collision time is
\begin{equation}
p \approx \mu \sigma_z = \frac{\alpha^2\sigma_z}{r_e \gamma} \Upsilon
T( \Upsilon )\;\;.
\end{equation}
From these equations we can find $\Upsilon$ for a certain conversion
probability $p$ (with an accuracy higher than 25\%)\cite{TEL90}
\begin{equation}
 \Upsilon_m = 2.7/ln(0.1/p_1) \hspace{22mm} p_1 < 0.01,
\end{equation}
$\hspace*{4.3cm} 1.2+9p_1    \hspace{18mm} 0.01< p_1 < 4,$ 

\vspace{0.4cm}

$\hspace*{3.8cm} 4.5\; p_1^{3/2}    \hspace{34mm} p_1 > 4, $ \\

where $$p_1=p\frac{r_e\gamma}{\alpha^2\sigma_z} \sim p\cdot 
0.1\frac{E[\TEV\ ]}{\sigma_z[\MM]}.$$

For the conversion probability $p$ the `geometrical` \GG\ luminosity is
suppressed approximately by a factor $e^{-p}$. 

We will see that maximum \GG\ luminosity is achieved at $p>1$. For $E
= 0.5-2.5\;$TeV and $\sigma_z = 0.1-0.5\;$ mm the parameter $p_1$
belongs to the second range (`transition' regime) where $\Upsilon \sim
1.2+9p_1$. 

\section{Estimation of  ultimate \GG\ luminosity}
  Let us find now limits posed on the luminosity due to coherent pair
creation for different collision schemes.

\
 There are three ways to avoid this effect (i.e. to keep
$\Upsilon \le \Upsilon_m$):

1) to use flat beams;

2) to deflect the electron beam after conversion at a sufficiently large
distance ($x_0$ for $E = E_0$) from the interaction point(IP);

3) under certain conditions (low beam energy, long bunches) $\Upsilon
< \Upsilon_m $ at the IP due to the repulsion of  electron beams
\cite{TELSH}.

Let us  consider at first requirements to the beam sizes in the case 1.

\subsection{Flat beams}

The field of the beam with the r.m.s horizontal size $\sigma_x$ and
the length $\sigma_z$ is $B\equiv|B|+|E| \sim
2eN/\sigma_x\sigma_z$. From the condition $\kappa \sim 0.8\gamma B/B_0 <
\Upsilon_m$ we get
\begin{equation}
\sigma_x > \frac{1.6N\gamma r_e^2}{\alpha\sigma_z\Upsilon_m} =
 \frac{1.6N\gamma
 r_e^2}{\alpha\sigma_z(1.2+9pr_e\gamma/\alpha^2\sigma_z)} \sim
 \frac{40\cdot\left[\frac{N}{10^{10}}\right]}
 {p+1.3\frac{\sigma_z[\MM]}{E[\TEV]}}\;\;\NM\
\end{equation} 
   The \GG\ luminosity at $z>0.65$
\begin{equation}
\LGG\ \sim \frac{0.5k^2N^2f}{4\pi(b/\gamma)\sigma_x}\sim 
\frac{0.025\alpha N \sigma_z f k^2}{br_e^2}
\left[1.2+9p\frac{r_e\gamma}{\alpha^2\sigma_z}\right]e^{-p},
\end{equation}
where the coefficient 0.5 follows from the simulation for
$\sigma_y=b\gamma$.  It has its maximum at
 
$$ I:\;\;\;\tilde{p}=0\;\;\;\;\mbox{at}\;\;
 a=\frac{7.5r_e\gamma}{\alpha^2\sigma_z} =
 \frac{0.75E[TeV]}{\sigma_z[\MM]} < 1\;\; ;$$

$$\hspace*{-3.6cm}II:\;\;\;\tilde{p}=1-1/a\;\;\; \mbox{at}\;\;
 a > 1\;\; .$$

The corresponding luminosities for these two cases are the following
\begin{equation}
\LGG\sim 0.03\frac{\alpha k^2Nf\sigma_z}{br_e^2}=
2.8\cdot 10^{33}\left(\frac{N}{10^{10}}\right)
\frac{f[kHz]}{b[\CM]}k^2\sigma_z[\MM],\; \CMS;
\end{equation}
\begin{equation}
\LGG\sim 0.23\frac{Nf\gamma k^2}{\alpha br_e}e^{-\tilde{p}}=
2.2\cdot 10^{33} \left(\frac{N}{10^{10}}\right)
\frac{f[kHz]}{b[\CM]}k^2 E[\TEV]e^{-\tilde{p}},\; \CMS.
\end{equation}

Optimum horizontal beam sizes in these two cases are
\begin{equation}
I:\;\;\; \sigma_x \sim \frac{1.3Nr_e^2\gamma}{\alpha\sigma_z} =
28 \frac{E[\TEV]\left(\frac{N}{10^{10}}\right)}{\sigma_z[\MM]},\;\NM\ ;\;
\mbox{at}\;\; a < 1\;\; ;
\end{equation}
\begin{equation}
\hspace*{-1cm}II:\;\;\; \sigma_x \sim 0.18\alpha N r_e = 37
\left(\frac{N}{10^{10}}\right),\; \NM\ 
\;\; \mbox{at}\;\; a > 1\;.
\end{equation}

  The minimum value of the distance between the conversion (CP)) and the
interaction regions $b$ is determined by the length of the conversion
region which is equal approximately to $b=0.08E[\TEV]$, cm (see section
6.1). For further estimation we assume that
\begin{equation}
b=3\sigma_z + 0.04E[\TEV], \;\;\CM.
\end{equation}
Let us take $N=10^{10},\;\sigma_z=0.2\;\MM,\;f=10\;kHz,\;k^2=0.4\;$
(one conversion length) that corresponds at $E>0.25\;$ \TEV\ to the
case II.  For $2E = 5$ TeV we get
\begin{equation}
\LGG \sim 6\cdot 10^{34}\; \CMS\ \;\;\mbox{at}\;\sigma_x \sim 40\;
\NM\ \;\; \mbox{and}\;\; \sigma_y \sim b/\gamma=0.3\;\NM. 
\end{equation}

For a very high energy $L_{max}\sim 8\cdot10^{34}\;\CMS\ $ for a chosen
parameters corresponding to the beam power $P = 15E[\TEV]\;$MW
per beam. In the next section we will compare these approximate
results with the results of simulation. 

\subsection{Influence of the beam-beam  repulsion on the coherent pair
 creation}

During the beam collision electrons get displacement in the field of the
opposing beam
\begin{equation}
r \sim \sqrt{\frac{\sigma_zr_eN}{8\gamma}}.
\end{equation}
This estimate is obtained from the condition that at the impact parameter
equal to the characteristic displacement the additional displacement
is equal to the initial impact parameter.

The field at the axis (which influences on the high energy  photons)
$B \sim 2eN/r\sigma_z$. Then the corresponding field parameter
\begin{equation}
\Upsilon \sim \gamma\frac{B}{B_0} = \frac{\gamma Br_e^2}{\alpha e} \sim
 5\frac{r_e\gamma}{\alpha\sigma_z}\sqrt{\frac{\gamma r_eN}{\sigma_z}}
\end{equation}

According to eq.(6), in the transition regime $\Upsilon_m = 1.2 +
9pr_e\gamma/\alpha^2\sigma_z$. From $\Upsilon = \Upsilon_m$ we can
find the maximum beam energy when the  coherent pair creation is
suppressed due to the beam repulsion. 

 At the energy $E>1$ TeV and bunches short enough, one can neglect the
first term and get
\begin{equation}
\gamma_{max}\sim 3\frac{p^2\sigma_z}{\alpha^2r_e N}\;\;\mbox{or} 
\;\;E_{max} \sim p^2\frac{\sigma_z[\MM]}{N/10^{10}},\;\TEV\ .
\end{equation}
The \GG\  luminosity is equal
\begin{equation}
\LGG(z>0.65) \sim 0.35\frac{N^2fk^2}{4\pi(b/\gamma)^2}e^{-p} \sim 
0.1(Nf)\frac{\sigma_z\gamma p^2 k^2}{\alpha^2 r_e b^2}e^{-p}, 
\end{equation}
where the numerical factor 0.35 follows from the simulation.
It has its maximum  at p=2 when
\begin{equation}
\LGG\sim 0.05(Nf)\frac{\sigma_z\gamma k^2}{\alpha^2 r_e b^2}\sim
7\cdot10^{33}\left(\frac{N}{10^{10}}\right)
\frac{\sigma_z[\MM]}{b^2[\CM]}E[\TEV]f[\mbox{kHz}]k^2\sigma_z[\MM].
\end{equation}
We have separated the factor (Nf) because it is a beam power. Taking in the
previous example $Nf = 10^{14}\;$ Hz, $\sigma_z =
0.2\;\MM,\;k^2=0.4,\;b=3\sigma_z+ 0.04E[\TEV],\;\CM,\;E=2.5\;\TEV\ $
we obtain
\begin{equation}
\LGG(z>0.65) \sim 6\cdot10^{35}\;\CMS.
\end{equation}
The optimum number of particles in the beam for an energy considered
(eq.(17)) is $N\sim0.8\cdot10^{10}.$

These estimates show that the beam repulsion substantially influences
(increases) the attainable \GG\ luminosity. This prediction will be
checked by the simulation.

  In fact, at low enough energy this effect allows to use even
infinitely narrow electron beams with any reasonable number of
particles in the bunch and the minimum photon spot size is $b/\gamma$,
where $b$ should be taken as small as possible. At high energies this
effect also works but the number of particles in the bunch should be
below some number dependent of the energy (eq.17).

\section{Scheme with magnetic deflection}
   Using the magnetic field $B_e$ between the conversion and interaction
regions one can sweep out electrons from the interaction point at some
distance $x_0$ sufficient for the suppression of coherent pair creation,
i.e. to satisfy condition $\Upsilon < \Upsilon_m$ given by eq.6. This
distance $x_0$ is  approximately equalto $\sigma_x$ given by
eq.7. The required distance $b$ is found from relation $x_0 \sim
b^2/2R = b^2eB_e/2E$. The photon spot size at IP is $b/\gamma$ and the
luminosity of \GG\ collisions 
\begin{equation}
\LGG(z>0.65) \sim 0.35\frac{k^2N^2f}{4\pi(b/\gamma)^2} \sim
\frac{0.03\alpha k^2 Nf
\sigma_zB_e}{er_e}(1+7.5p\frac{r_e\gamma}{\alpha^2\sigma_z})e^{-p}
\end{equation}

   The optimization over $p$ gives 

$$\LGG\sim \frac{0.03\alpha k^2 Nf\sigma_zB_e}{er_e} = $$
\begin{equation}
= 1.6\cdot10^{34}\left(\frac{N}{10^{10}}\right)\sigma_z[\MM]f[kHz]B_e[T]k^2
\;\;\mbox{at}\;\; a= \frac{0.75E[TeV]}{\sigma_z[\MM]} < 1\; ;
\end{equation}

$$\LGG\ \sim \frac{0.22 k^2 Nf\gamma B_e}{\alpha e}e^{-(1-1/a)} = $$
\begin{equation}
 = 1.25\cdot 10^{34}\left(\frac{N}{10^{10}}\right)
f[kHz]B[T]E[\TEV]k^2 e^{-(1-1/a)},\; \CMS;
\;\;\mbox{at} \;a>1\;.
\end{equation}
 As before, taking $N=10^{10},\;f=10\;\mbox{kHz},
\;\sigma_z=0.2\;\MM,\;\;k^2=0.4\;E=2.5\;\TEV\ $ and $B_e=0.5\;$T we get
\begin{equation}
\LGG(z>0.65)\sim 2.5\cdot 10^{34}\;\; \CMS.
\end{equation}

This number is notably smaller than that in the scheme without deflection
eq(20). The luminosity is proportional to $B_e$ and one can take
larger field values  but it poses some  technical problems. One should
also remember that in the transverse magnetic field, the soft background
particles produced in the forward direction (mainly \EPEM\ pairs) get
a kick, begin to spiral in the detector field and to avoid
backgrounds the radius of the vacuum pipe should exceed
$r=2bB_{\perp}/B_{\parallel}$.

  Considering the scheme with deflection we did not consider the field
created by the produced \EPEM\ pairs. These particles are closer to the
beam axis than the deflected beam and their field can even exceed the
field of the opposing electron beam. Therefore, our optimization of
pair creation  overestimates the luminosity. 
 We will see results of simulation in next sections. 

\section{Simulation}
\subsection{Assumptions}

We have seen that the picture of collisions is very complicated. It is
easier to get the result by simulation. The simulation code used in
this work~\cite{TEL95} includes all important processes. In the
present study, the beams are collided in very ultimate conditions:
very small beam sizes, high energies, too many beamstrahlung
photons. In order to avoid the time consumption problem only one
simplification was done: charged particles emitted the beamstrahlung
photons during the beam collision but these photons were excluded from
further consideration. It was assumed that:

a) the thickness of the laser target is equal to one collision length
($k=1-e^{-1}\sim 0.6$);

b) electrons and laser photons are polarized and  $2P_c\lambda_e = -1$;

c) varing the number of particles in the beam we kept constant beam power
$NfE=15E[\TEV],\;$MW;

d) the minimum distance between the CP and IP region was taken to be
 $b=0.2+0.1E[\TEV],\;\CM$ for fig.~\ref{sb1} and $b=3\sigma_z +
 0.04E[\TEV]$ \CM\ for the rest figures. In the later case, the
 parameter $\xi^2$ characterizing the nonlinear effect in the Compton
 scattering~\cite{TEL95} is equal to 0.6;

e) the vertical beam size is equal to $0.5b/\gamma$;

\subsection{Simulation results}

\begin{figure}[!hb]
\centering
\epsfig{file=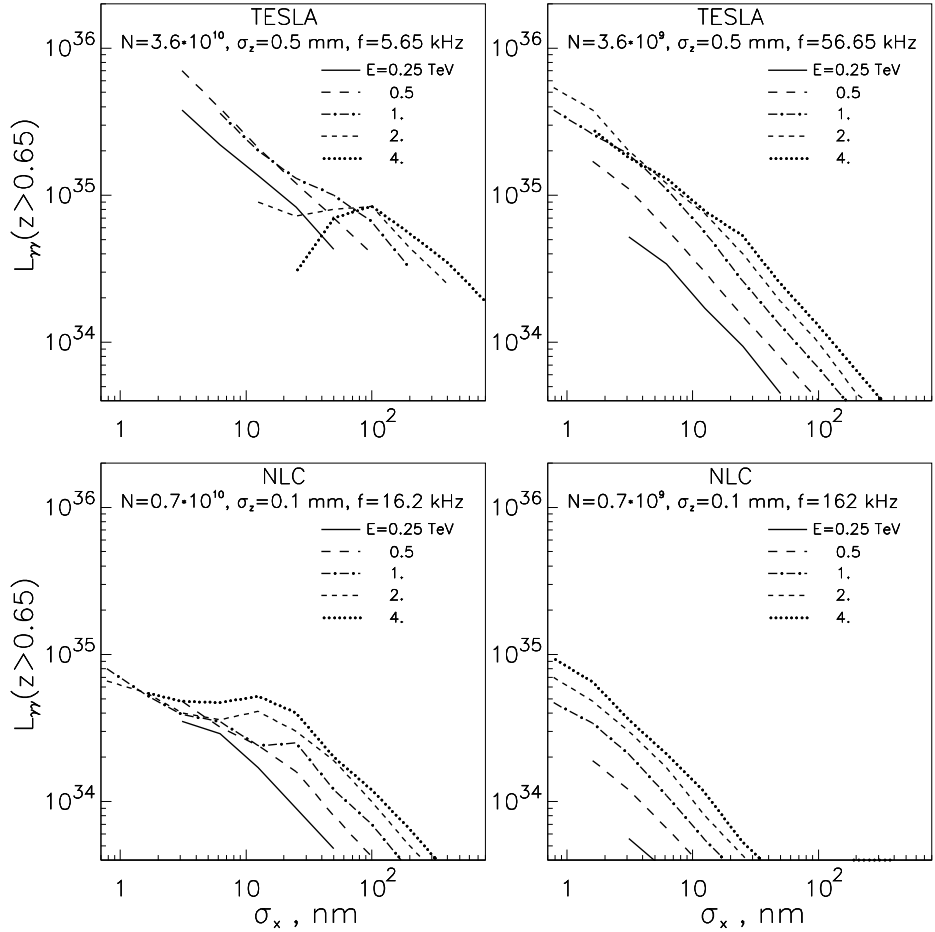, height=14.5cm,angle=0}
\caption{Dependence of the \GG\ luminosity on the horizontal beam size
for TESLA and NLC beam parameters, see comments in the text.}
\label{sb1}
\end{figure}

 Fig.~\ref{sb1} shows \GG\ luminosity as a function of $\sigma_x$ at
 TESLA and NLC for the beam energy range 0.25--4 TeV. On the righthand
 graphs, the number of particles was decreased by a factor 10, while
 the collision rate was increased by the same factor. The distance
 $b=0.2 + 0.1E[\TEV]$ \CM\ . We see on the left-hand side graphs that
 for `nominal' numbers of particles in the beams the luminosity does
 not follow the dependence $L \propto 1/\sigma_x$ due to the
 conversion of photons to \EPEM\ pairs. It happens at $\sigma_x$ very
 close to our prediction, eq(12).

\begin{figure}[!hb]
\centering
\epsfig{file=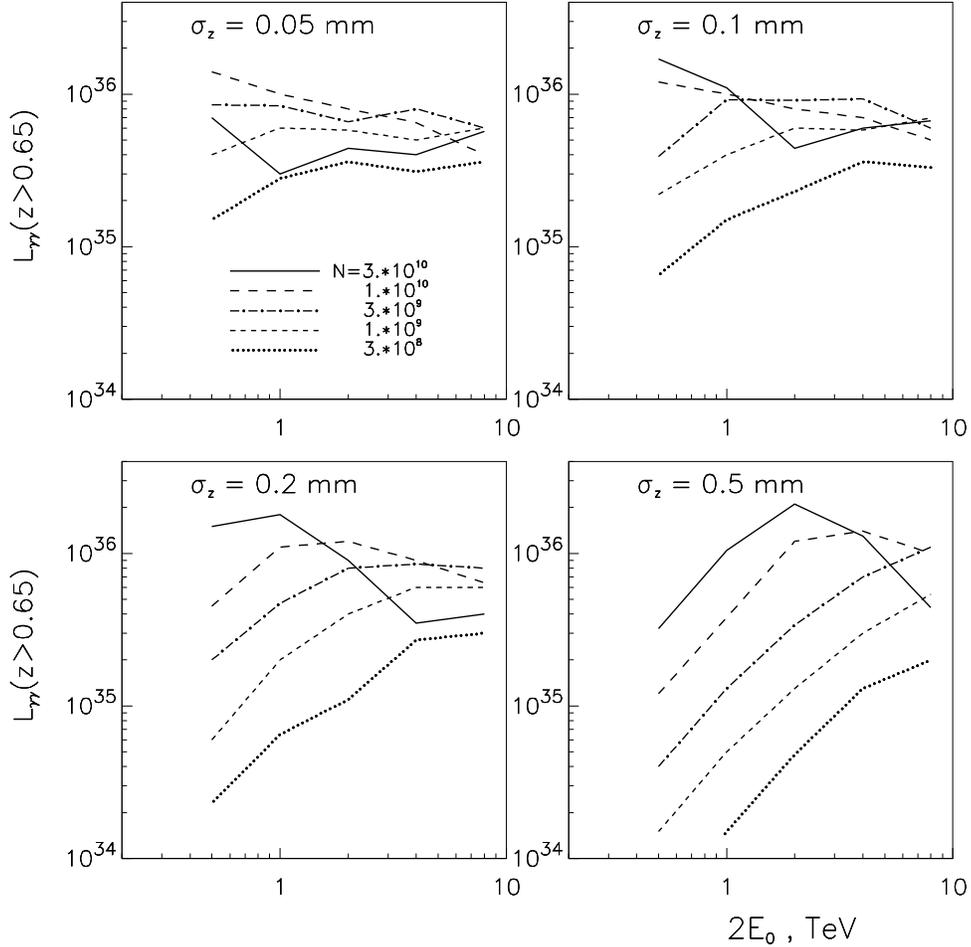, height=14.5cm,angle=0}
\caption{\GG\ luminosity for round beams at the minimum distance between
interaction region and collision points, see comments in the text.}
\label{sb2}
\end{figure}

In fig.~\ref{sb2} we can see the dependence of the luminosity both on N and
 $\sigma_z$ in the case where beams are round and the conversion
 region is situated as close as possible: $b=3\sigma_z + 0.04E[\TEV]$
 \CM\ . The total beam power is 15$E[\TEV]$ MW. Looking to this
 pictures one can make many own observations which are clear after our
 theoretical consideration. Note only that the longer bunch requires
 the larger laser flash energy for conversion ($A\propto \sigma_z$) and
 not every linac (among the current projects) can accelerate a 0.5 mm long
 bunch due to wake fields.  Let us take for further study
 $\sigma_z=0.2$ mm.

\begin{figure}[!hb]
\centering
\epsfig{file=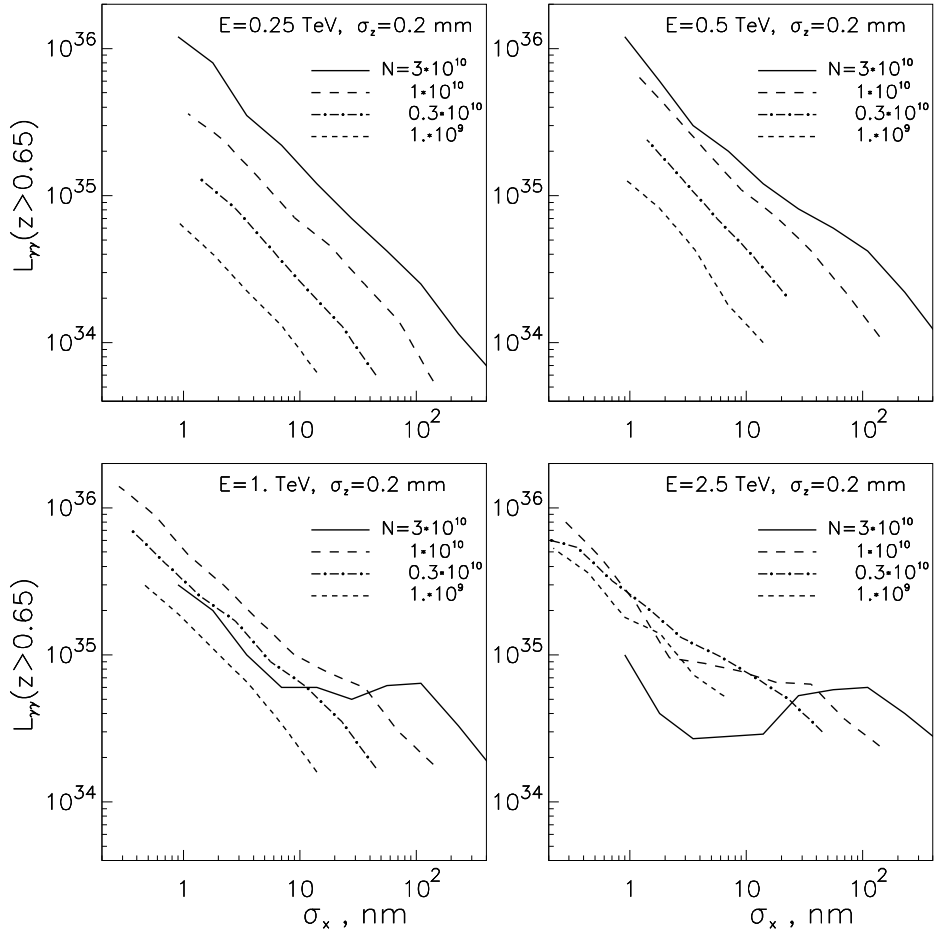, height=14.5cm,angle=0}
\caption{Dependence of the \GG\ luminosity on the horizontal beam size
for $\sigma_z = 0.2$ \MM, see comments in the text.}
\label{sb3}
\end{figure}

  The dependence of the luminosity on $\sigma_x$ (other conditions are
  the same as in the previous figure) is depicted in
  fig.~\ref{sb3}. It is desirable to choose the working point
  ($\sigma_x,\;N$) so that not only the luminosity is large but also
  the corresponding curve still follows their natural behavior $L
  \propto 1/\sigma_x$. This means that the $\gamma \to \EPEM\ $
  conversion probability is still not too high. Otherwise it may
  happen that the low energy \GG\ luminosity will be much larger than
  that in the high energy part (low energy photons have smaller
  probability of conversion). For example: at E = 2.5 TeV one can
  reach $\LGG \sim 7\cdot10^{34}\;\CMS\ $with all considered number of
  particles in the bunch.

  The values of $\sigma_x$ and \LGG\ where curves make zigzag are in
  good agreement with eqs. (10) and (12). Even higher luminosities at
  E = 2.5 TeV can be reached with $N=10^9$--$10^{10}$, but in the case
  $N=10^{10}$ it happens after zigzag on the curve that manifests that
  many photons have converted to \EPEM\ pairs. The best choice here is
  $N=0.3\cdot10^{10}$, where one can reach $L=4\cdot10^{35}$ without
  problems.  The only problem here is a too small transverse beam size:
  $\sigma_x,\sigma_y < 1\; \NM$. 

  At low energies the situation is perfect
and one can dream about WW factory at the collider with an energy
2E = 500--1000 GeV and luminosity upto $\LGG\ \sim 10^{35}\;\CMS.$

\begin{figure}[!hb]
\centering
\epsfig{file=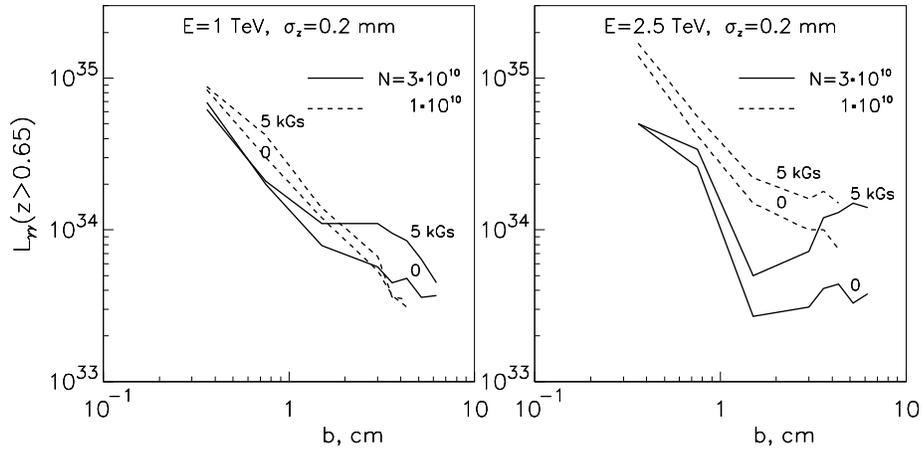, height=8 cm,angle=0}
\caption{Dependence of the \GG\ luminosity on the distance between
interaction point and conversion region in the scheme with magnetic
deflection, see comments in the text.}
\label{sb4}
\end{figure}

Fig.~\ref{sb4} presents the result for the scheme with magnetic deflection. 
The strength of magnetic field is equal 0.5 T and the distance $b$ is
varied. For the comparison the case $B_e=0$ is also shown. We see that
the magnetic deflection helps at large $b$ but the luminosity value
in this region is much lower than that in the case of flat beams where
conversion point is situated very close to the IP. Nevertheless,
although magnetic deflection does not help to reach ultimate
luminosities, in `practical` cases (where the luminosity is far from the
limit) the magnetic deflection helps to
decrease the low energy \GG\ luminosity without degradation of the high
energy part.

Comparing our luminosity scaling law (eq.1) with the results of the
simulation we see that the required luminosity can be reached at least
up to $2E= 4\;\TEV\ $ with any (up to $3\cdot10^{10}$) number of
particles in the bunch. Note that in our examples the collision rate
is $f$=10 kHz$(10^{10}/N)$. The higher $f$ is good for the experiment
but makes some problems for the required average laser power. It seems
that f=10--30 kHz is still acceptable. With N=0.3$\cdot10^{10}$ one can
reach at 2E=5 TeV even a few times higher luminosity than that it is
`required'. The main problem here is connected with attainable beam
emittances.

\section{Beam emittances}

 Only short remark. The normalized emittances which are written in the
 current projects \cite{LOW} do not allow to follow the scaling low
 above 2E = 1 TeV. An increase in luminosity at low energies or motion
 towards high energies requires serious R\&D work on low emittance
 electron beams. One new suggestion was reported at the ITP
 Workshops\cite{TCOOL}. This is laser cooling which allows to cool
 beams down to values necessary for all  parameters of
 photon colliders considered here.

\section{Conclusion}

At $2E\sim 5 \;\TEV\ $ the conventional linear colliders reach their limit
both in \EPEM\ and in \GG\ mode. The number of problems grows
exponentially: acceleration gradient, very small beam sizes,
radiation, pair creation. The reason is common for all mode of operation:

1) the required luminosity is proportional to $E^2$; 

2) energetic problem,  because beams are used only once (but namely this
feature makes possible to consider photon colliders).

Linear colliders are perfect for 2E = 0.1 -- 2 TeV and we have to use their
potential with highest efficiency. 

\section*{Acknowledgments} 
I would like to thank Z.Parza, the organizer of the
Program ''New Ideas for Particle Accelerator'' at ITP, UCSB, Santa
Barbara, supported with National Science Foundation Grant NO
PHY94--07194.

\vspace*{0.5cm}

\end{document}